\documentclass[hyper]{JHEP} 
\title{Proposal for Background Independent
 Berkovits'Superstring Field
Theory}  \author{by J. Kluso\v{n}\\ 	 Department of Theoretical Physics and Astrophysics\\                    Faculty of Science, Masaryk University\\ Kotl\'{a}\v{r}sk\'{a} 2, 611 37, Brno\\ Czech Republic\\ 	E-mail: \email{klu@physics.muni.cz}}   \preprint{\hepth{0106107}}  			  	
 \abstract{In this paper we would like to propose the
background independent formulation of Berkovits' superstring
field theory. Then we will show that  the solution of equation
of motion of this theory leads to the Berkovits' superstring
field theory formulated around particular
CFT background.}


\begin{document}
\section{Introduction}\label{first}

It was shown in \cite{Horowitz} that the
fundamental  formulation
of bosonic open string field theory
\cite{WittenSFT} can be formulated as a
pure cubic string field theory action. This action does not
contain any BRST operator and it is
formally background independent. When
we  then expand string field around any solution
of the equation of motion arising from this
pure cubic string field   action
 we  obtain
exactly Witten's bosonic field theory with
the BRST operator constructed  from
the field obeying the equation of motion of
the pure cubic string field theory. This approach
has been further developed in \cite{Horowitz1},
where another solutions of the pure cubic string field
theory that do not correspond to any usual BRST
operator were found. 

Recently there was a great interest in 
the formulation of the open bosonic string
field theory around  the closed string vacuum
(Vacuum string field theory-VSFT)
\cite{SenV1,Hata,Taylor,Feng,Feng1,SenV2,SenV3,
Gross1,SenV4,Muki,David,SenV5,Gross2} that
arises as a final point of the tachyon condensation
on  D25-brane (For a very nice review of the string
field theory and its relation to the problem of the tachyon
condensation, see \cite{Ohmori}.) VSFT is
characterised by   BRST operator with 
trivial cohomology so that there are not any
physical open string excitations. This BRST
operator is constructed
from ghost fields only so that vacuum string field theory
is formally background independent.  VSFT  seems
 to be a very promising area of research which 
could lead to better understanding of the
string theory. 

The problem of the tachyon condensation was
studied in superstring theory as well. It seems 
that the most promising approach to the study
of the tachyon condensation in this context is based
on Berkovits' superstring field theory
\footnote{For another interesting formulation
of superstring field theory, see \cite{Zubarev1,Zubarev2,Zubarev3}.}
\cite{Berkovits2,Berkovits3}.
Two major advantages
of Berkovits superstring field theory
 are that it is manifestly
$SO(3,1)$ super-Poincare invariant and
that it does not require contact terms to
remove tree-level divergences
\cite{BerkovitsD} (For recent
review of Berkovits superstring field theory 
(NSFT), see \cite{BerkovitsR}.)
In particular, it was shown that the calculation
of the tachyon potential in NSFT is in a
very good agreement with Sen's conjecture
\cite{SenP}.
The tachyon kink and lump solution was
also analysed (For recent work, see \cite{Ohmori1},
for list of references, see \cite{Ohmori}.) However
this theory is formulated for Neveu-Schwarz open
string sector only because it is not known how
to include Ramond sector in manifestly Lorentz invariant
manner.

In this paper we would like to ask  the question 
whether there could be such a formulation of
 string field theory
from which the NSFT arises in a natural way
 as the Witten's open string field
theory emerges from the pure cubic string field action.
Since  pure cubic string field theory does not contain
any BRST operator it is formally background
independent,  we can ask the 
question whether there is  such a formulation of the
NSFT theory that   is   background independent as well.
In other words, we are looking for a string field theory action
that does not depend on any particular matter conformal 
field theory background. It seems to us
 that the second
formulation of the problem is the appropriate one
for Berkovits' superstring field theory.
  This can be seen as follows. 
In Witten's string field theory the fundamental
object is a ghost number one string field. Since the BRST
operator carries the same ghost number there is no problem 
to define BRST operator using string field as has been
done in \cite{Horowitz}. On the other hand, fundamental
string field in NSFT is a string field of the ghost number
and the picture number zero. At first sight it seems to
be impossible to define string field theory action 
in background independent way. However recent
results  \cite{SenV1,SenV2,SenV3,SenV4} suggest
that there can exist BRST operator that is constructed
purely from the ghost field  which is universal for any
conformal field theory background. In fact, in a remarkable
paper \cite{BerkovitsBRST} the BRST operator
 for $N=1$ NSR string theory was given
that resembles striking similarity with the terms
presented in NSFT action
\footnote{We are very thankful to
N. Berkovits for emphasising this form of the
 BRST operator and pointing out
the reference \cite{BerkovitsBRST}.}
. It is then natural to presume
that the background independent formulation of the
NSFT theory could be based on the BRST operator $Q_0$
defined in \cite{BerkovitsBRST} and  explicit
form of which will be given below
\footnote{Similar suggestion has been  proposed in
\cite{Berkovitsunpub}.}. Then using  recent result
\cite{KlusonNSFT} it is possible to find such a
string field  that is solution of the equation of motion of the
background independent NSFT and 
that leads to the emergence of the correct  BRST operator
for any CFT background. 

The plan of the paper is follows. In the
next section (\ref{second}) we briefly review 
NSFT theory defined on the BPS D-brane. 

In section (\ref{third}) we  review the
 formulation of the open bosonic string
field theory based on the pure cubic action
\cite{Horowitz,Horowitz1}. We will show how
Witten's open bosonic string field theory action
naturally emerges from the pure cubic string
field action. We use these elegant calculations
 in  section
(\ref{fourth}) where we propose a background
independent formulation of the Berkovits string
field theory. Then we will show that from
this action we can obtain any NSFT action
for any CFT background.

In conclusion (\ref{fifth}) we will outline our
 results  and suggest
further extension of this work.

\section{Review of superstring field theory}
\label{second}
 
In this section we would like to review basic facts about
Berkovits superstring field theory, for more details, see \cite{Ohmori,
Berkovits1,Berkovits2,Sen1,BerkovitsR}. 
The general off-shell string field configuration
in the 
\footnote{In this paper we will consider 
NSFT for BPS D-branes only.}
GSO(+) NS sector corresponding Grassmann
even open string vertex operator $\Phi$ of ghost number
$0$ and picture number $0$ in the combined
conformal field theory of a $c=15$ superconformal
matter system, and $b,c,\beta,\gamma$ ghost
system with $c=-15$. We can also express $\beta,
\gamma$ in terms of ghost fields
$\xi,\eta,\phi$
\begin{equation}
\beta=e^{-\phi}\partial \xi , \
\gamma=\eta e^{\phi} \ ,
\end{equation}
the ghost number $n_g$ and the picture number $n_p$
assignments are as follows
\begin{eqnarray}
b: \ n_g=-1,\ n_p=0 \ \ \ c: \ n_g=1, \ n_p=0 \ ; \nonumber \\
e^{q\phi}: \  n_g=0, \ n_p=q \ ; \nonumber \\
\xi: \ n_q=-1 , \ n_p=1 \ \ \ \eta: \ n_q=1, \ n_p=-1 \ . \nonumber \\
\end{eqnarray}
The BRST operator $Q_B$ is given
\begin{equation}
Q_B=\oint dz j(z)=\int dz
\left\{c(T_m+T_{\xi\eta}+T_{\phi})+c\partial cb+\eta e^{\phi} G_m
-\eta \partial \eta e^{2\phi}b\right\} \ ,
\end{equation}
where
\begin{equation}
T_{\xi\eta}=\partial \xi \eta, \
T_{\phi}=-\frac{1}{2}\partial \phi\partial\phi-\partial^2\phi \ ,
\end{equation}
$T_{m}$ is a matter stress tensor and $G_m$ is a matter superconformal 
generator. Throughout this paper we will be working in
units $\alpha'=1$.

The string field action is given \cite{Berkovits1,Berkovits2}
\begin{equation}\label{NSFTaction}
S=\frac{1}{2}\int\left(
(e^{-\Phi}\star Q_Be^{\Phi})(e^{-\Phi}\star\eta_0
e^{\Phi})-\int_0^1 dt
(e^{-t\Phi}\star\partial_te^{t\Phi})\star\left\{
(e^{-t\Phi}\star Q_Be^{t\Phi}),
(e^{-t\Phi}\star\eta_0e^{t\Phi})\right\}\right) \ ,
\end{equation}
where $\{A,B\}=A\star B+B\star A$ and $e^{-t\Phi}\star\partial_t
e^{t\Phi}=\Phi$. Here the products and
integral are defined by Witten's gluing prescription
of the string. The exponential of string field is
defined in the same manner $e^{\Phi}=
1+\Phi+\frac{1}{2}\Phi\star\Phi+\dots$. 
The basis properties of $Q_B,\eta_0$ which we
will need in our analysis (for more details, see
\cite{Ohmori} and reference therein) are
\begin{eqnarray}\label{ax}
Q_B^2=0, \ \eta_0^2=0, \ \{Q_B,\eta_0\}=0 \ ,
\nonumber \\
Q_B(\Phi_1\star\Phi_2)=Q_B(\Phi_1)\star\Phi_2+
\Phi_1\star Q_B(\Phi_2) , \ \nonumber \\
\eta_0(\Phi_1\star\Phi_2)=\eta_0(\Phi_1)\star\Phi_2+
\Phi_1\star\eta_0(\Phi_2) , \ \nonumber \\
\int Q_B(\dots)=0 \ , \int \eta_0(\dots)=0 \ , \nonumber \\
\end{eqnarray}
where $\Phi_1,\Phi_2$ are Grassmann even fields. 

For our purposes it will be useful to write the BRST operator
in the form \cite{BerkovitsBRST} 
\begin{equation}\label{BRSTR}
Q_B=e^{-R}\left(
\frac{1}{2\pi i}\oint dz \gamma^2 b
\right)e^R \ ,
\end{equation}
where
\begin{equation}
R=\frac{1}{2\pi i}\oint dz [c G_m e^{-\phi}e^{\chi}-
\frac{1}{4}\partial (e^{-2\phi})e^{2\chi}c\partial c ]=
\frac{1}{2\pi i}\oint dz r(z) \ 
\end{equation}
and
\begin{eqnarray}
\beta=e^{-\phi}\partial \xi \ , \gamma=\eta e^{\phi}, \
\xi(z)\eta(w)=\frac{1}{z-w}, \ \phi(z)\phi(w)=-\log (z-w) \ ,
\nonumber \\
\xi=e^{\chi}, \eta=e^{-\chi}, \  
\chi(z)\chi(w)=-\log(z-w) \ . \nonumber \\ 
\end{eqnarray}
It is easy to see that
\begin{equation}\label{BRST0}
Q_0^2=\frac{1}{2}\{Q_0,Q_0\}=
0  \ , Q_0=\frac{1}{2\pi i}\oint \gamma^2 b \ 
\end{equation}
and also
\begin{equation}
\{\eta_0,Q_0\}=0 \ .
\end{equation}
Then we immediately obtain
\begin{equation}
Q_B^2=e^{-R}Q_0^2e^{R}=0 \ .
\end{equation}
In other words, $Q_B$ is a nilpotent operator.
It was also shown in \cite{BerkovitsBRST} that
$Q_B$ given in (\ref{BRSTR}) anticommutes with
$\eta_0$ in critical dimensions $D=10$.
 From the fact that $Q_0$ is
constructed from the ghost fields only it is the
same for any CFT background so that it seems to be
a  natural BRST operator 
for a background  independent formulation of NSFT 
in the similar way as in \cite{SenV1,SenV2,SenV3,SenV4}.
On the other hand, it must be stressed that this
operator has not  trivial cohomology 
\footnote{We would like to thank N. Berkovits for
stressing this point to us.} so that the background
independent formulation of NSFT should not be confused with
the vacuum string field theory that  describes the
bosonic open string field theory
 after the tachyon condensation. We mean that this 
seems to be natural fact since we are looking for a background
independent formulation of the BPS D-brane so 
that there is not any tachyon field present and hence
the process of the tachyon condensation cannot occur.

In the next section we briefly review the approach
given in \cite{Horowitz}. In particular, we will see how
Witten's open bosonic string field theory naturally emerges
from the pure cubic string field theory action. 
\section{Pure Cubic Action for String Field
Theory}\label{third}
It was proposed in \cite{Horowitz} that pregeometrical,
background independent (at least formally) string
field action has a form
\begin{equation}\label{Cubicaction}
S=\frac{1}{3}\int \Phi\star\Phi\star \Phi \ ,
\end{equation}
where $\Phi$ is  Grassmann odd string field of ghost
number one. The classical equation of motion is
\begin{equation}
\Phi\star\Phi=0 \ .
\end{equation}
As was shown in \cite{Horowitz} when we expand around
the classical solution 
\begin{equation}
\Phi=\Phi_0+\phi
\end{equation}
we  get an action for fluctuation
\begin{equation}
S=\int \frac{1}{2}\phi\star D_{\Phi_0}\phi+
\frac{1}{3}\phi\star\phi\star\phi \ ,
\end{equation}
where
\begin{equation}
D_{\Phi_0}X=\Phi_0\star X-(-1)^XX\star \Phi_0 \ .
\end{equation}
It was shown in \cite{Horowitz} that $D_{\Phi_0}$
is a derivation. In order to recovery Witten's
form of the string field action \cite{WittenSFT} we must find field
$\Phi_0$ solving equation of motion and
\begin{equation}
D_{\Phi_0}=Q_B \ ,
\end{equation}
where $Q_B$ is the BRST operator associated with
some background. It was shown in
\cite{Horowitz} that such a field $\Phi_0$
is uniquely given  by the relation
\begin{equation}
\Phi_0=Q^L\mathcal{I} \ ,
\end{equation}
where $Q^L \ (Q^R)$ is the BRST charge density
integrated over the left (right) half of the string
($Q=Q^R+Q^L$) and $\mathcal{I}$
is the identity operator of the algebra obeying
\begin{equation}
\mathcal{I}\star B=B\star \mathcal{I}=B, \  \forall B \ .
\end{equation}

In Witten's open string field theory, $Q_B$
represents a reference background and $\phi$
represents the second quantized fluctuation field
around that  background. As was shown in
\cite{Horowitz}, shifting $\phi$ it is possible to
eliminate this specific reference to a background.
In their second-quantized formulation the
backgrounds arise as solutions to the equations of
motion. 

It is natural to ask the question whether similar 
formalism works in case of NSFT as well. Although
we will not be able to find an analogue of the
pure cubic string field action for NSFT, 
 we will see  that it is possible
(at least formally) to formulate the background independent
version of NSFT based on the BRST operator
constructed from the ghost field only.

\section{Proposal for background independent
NSFT}\label{fourth}

Since we will not perform any explicit
calculation we will again use abstract
Witten's formalism in string field theory
\cite{WittenSFT}. 
We would like to find a formulation
of the NSFT theory from which a general action
(\ref{NSFTaction}) emerges in natural
way as in \cite{Horowitz,Horowitz1}, or equivalently,
we  
would like to find a formulation of
the NSFT theory that does not depend on
any particular  CFT background as in
case of VSFT. In fact, the second requirement is
the more appropriate one for NSFT which
can be seen from the following argument.
In bosonic string field theory the fundamental
object is a string field of the ghost number 
one so that it is natural that the BRST operator
of the same ghost number can emerge
from the pure cubic action. On the other
hand in NSFT theory the fundamental field
is a Grassmann ghost zero field so that it
does not seem to be possible construct BRST
operator from the action containing fundamental
fields only without presence of any ghost
number one operator. However, there is certainly
fundamental BRST operator that does not
depend on any particular background, which
is the  operator given in \cite{BerkovitsBRST}
\begin{equation}
Q_0=\frac{1}{2\pi i}\oint dz \gamma^2(z)b(z) \ .
\end{equation}
It is clear that this operator does not depend on
any background CFT so it is natural to propose
background independent formulation of
NSFT theory using this  nilpotent operator
\begin{equation}\label{VNSFT}
S=\frac{1}{2}\int\left(
(e^{-\Phi}\star Q_0e^{\Phi})(e^{-\Phi}\star\eta_0
e^{\Phi})-\int_0^1 dt
(e^{-t\Phi}\star\partial_te^{t\Phi})\star\left\{
(e^{-t\Phi}\star Q_0e^{t\Phi}),
(e^{-t\Phi}\star \eta_0e^{t\Phi})\right\}\right) \ .
\end{equation}
We would like to show, following recent analysis
\cite{KlusonNSFT} that it is possible to find
such a solution of the equation of motion
arising from the previous action
\begin{equation}\label{eqm1}
\eta_0(e^{-\Phi_0}\star Q_0(e^{\Phi_0}))=0 \ 
\end{equation}
that leads to the NSFT action with appropriate 
BRST operator $Q_B$. 

Let us consider any string field $\Phi_0$, corresponding to
$G_0=e^{\Phi_0}$,  which is a solution of the equation 
of motion (\ref{eqm1}). In order to find a new form
of BRST operator, we must study the behaviour of the
 fluctuation field 
around this  solution \cite{KlusonNSFT}. For that reason
we write general string field containing 
fluctuation around this solution as
\begin{equation}
G=G_0\star h, \  h=e^{\phi}, \ G^{-1}=h^{-1}\star G^{-1}_0 \ . 
\end{equation}
To see  that this  field really describes
 fluctuations around  solution $G_0$ note
that for $\phi=0, G=G_0$. It is also clear that
any  string field in the form $e^{\Phi_0+\phi'}$ can be
always rewritten in the form  given above.

  Inserting this upper 
expression in (\ref{VNSFT}) we obtain an action for
$\phi$. As was argued in \cite{KlusonNSFT} in order 
to find a new BRST operator
we must ask the question what form
of the equation of motion obeys shifted field $h=e^{\phi}$.
 Then it was shown that the new BRST operator has 
a form
\begin{equation}\label{newBRST}
Q_B(X)=Q_0(X)+A\star X-(-1)^XX\star A \ ,
A=G_0^{-1}\star Q_0(G_0) \ 
\end{equation}
and the string field action for the fluctuation field
has exactly the same form as (\ref{NSFTaction}) with
the BRST operator (\ref{newBRST}).

From (\ref{newBRST})  we obtain
\begin{equation}
(Q_B-Q_0)X=A\star X-(-1)^XX\star A \ .
\end{equation}
Using results given in \cite{Horowitz}
\begin{eqnarray}\label{Qhor}
Q^R\mathcal{I}=-Q^L\mathcal{I} \ , Q=Q^R+Q^L,  \ 
\mathcal{I}\star X=X\star\mathcal{I}=X, \  \forall X \ ,
\nonumber \\
(Q^R X)\star Y=-(-1)^XX\star Q^L(Y) , \forall X,Y \ , \nonumber \\
\{Q,Q^L\}=0 \ , \nonumber \\
\end{eqnarray}
we see that we can write $A$ as
\begin{equation}\label{A}
A=(Q_B-Q_0)^L\mathcal{I}
\end{equation}
since then
\begin{eqnarray}
(Q_B-Q_0)^L(\mathcal{I})\star X=
-(Q_B-Q_0)^R(\mathcal{I})\star X=
\mathcal{I}\star (Q_B-Q_0)^L(X) \ , \nonumber \\
-(-1)^XX\star (Q_B-Q_0)^L\mathcal{I}=
(Q_B-Q_0)^R(X)\star\mathcal{I} \ , \nonumber \\
\end{eqnarray}
after application of (\ref{Qhor}). Then we have
\begin{equation}
A\star X-(-1)^XX\star A=
(Q_B-Q_0)^L(X) +
(Q_B-Q_0)^R(X)=(Q_B-Q_0)(X) \ .
\end{equation}
Since we know that $Q_B,\ Q_0$ are correct
BRST operators that anticommute with $\eta_0$
\cite{BerkovitsBRST}, we
can easily prove that $A$ given in (\ref{A})
solves the equation of motion (\ref{eqm1}). This
can be seen as follows. The fact that $\{\eta_0,G_0\}=
\{\eta_0,Q_B\}$  means that  when we express 
these operators using appropriate currents then
OPE between $\eta(z)$ and $j_0(z),j_B(z)$ is non-singular
\begin{equation}
\eta (z)j_0(w)\sim O(0) \ , 
\eta(z)j_B(w) \sim O(0) \ ,
\end{equation}
If we do a contour   integral over $z$
 in upper expressions we obtain
the  operator $\eta_0$ and next integration of $j_{0,B}(w)$
 over left half of the string
 we  find that the anticommutator is
equal to zero
\begin{equation}
\{\eta_0,Q_0^L\}=0 \ ,
\{\eta_0,Q_B^L\}=0 \ .
\end{equation}
Consequently we get from (\ref{A}) 
\begin{equation}\label{A1}
\eta_0 (A)=\eta_0(Q_B-Q_0)^L\mathcal{I}=
-(Q_B-Q_0)^L(\eta_0(\mathcal{I}))=0 \ ,
\end{equation}
where we have used
\begin{equation}
\eta_0(X)=\eta_0(X\star \mathcal{I})=
\eta_0(X)\star \mathcal{I}+(-1)^XX\star\eta_0(\mathcal{I})
\Rightarrow \eta_0(\mathcal{I})=0 \ .
\end{equation}
From (\ref{A1}) we see that $A$ given in (\ref{A})
solves the equation of motion (\ref{eqm1}). Now we
explicitly determine $\Phi_0$ in $A=e^{-\Phi_0}Q_0(e^{\Phi_0})$.

Let us define the function
\begin{equation}
F(t)=e^{-t\Phi_0}Q_0(e^{t\Phi_0})
\end{equation}
We can make an expansion around the point $t=0$ where
$F(0)=Q_0(1)=0$ 
\begin{equation}
F(t)=\sum_{n=0}^{\infty}\frac{1}{n!}\left.
\frac{d^nF(t)}{dt^n}\right|_{t=0}
t^n \   .
\end{equation}
The first derivative is equal to
\begin{equation}
\frac{dF}{dt}=-e^{-t\Phi_0}\Phi_0 Q_0(e^{t\Phi_0})+
e^{-t\Phi_0}Q_0(\Phi_0 e^{t\Phi_0})=
e^{-t\Phi_0}Q_0(\Phi_0)e^{t\Phi_0} \ 
\end{equation}
and the second one 
\begin{equation}
\frac{d^2 F(t)}{dt^2}=-e^{-t\Phi_0}
\Phi_0 Q_0(\Phi_0)e^{t\Phi_0}+
e^{-t\Phi_0}Q_0(\Phi_0)\Phi_0 e^{t\Phi_0}
=e^{-t\Phi_0} 
[Q_0(\Phi_0),\Phi_0]e^{t\Phi_0} \ .
\end{equation}
Generally we have
\begin{equation}
\frac{d^n F(t)}{dt^n}=
e^{-t\Phi_0}
\overbrace{[[Q_0(\Phi_0),\Phi_0],\dots],\Phi_0]}^{n-1}
e^{t\Phi_0}, \ n>1 
\end{equation}
and consequently
\begin{equation}
F(t=1)=A=e^{-\Phi_0}Q_0(e^{\Phi_0})=
Q_0(\Phi_0)+\sum_{n=2}^{\infty}
\frac{1}{n!}\overbrace{[[Q_0(\Phi_0),\Phi_0],\dots,]\Phi_0]}^{n-1} \ .
\end{equation}
We would like to compare this expression with
the  expression for BRST operator 
\cite{BerkovitsBRST} given in (\ref{BRSTR})
and with $Q_0$ given in (\ref{BRST0}).

As in calculation performed above we define
\begin{equation}
F(t)=e^{-Rt}Q_0e^{Rt}=F(0)+\sum_{n=1}^{\infty}
\frac{1}{n!}\frac{d^n F(t)}{dt^n}t^n
\end{equation}
then
\begin{eqnarray}
F(0)=Q_0 , \
\frac{dF}{dt}=-e^{-tR}RQ_0e^{tR}+e^{-tR}Q_0Re^{tR}=
e^{-tR}[Q_0,R]e^{tR}, \ \nonumber \\
\frac{d^2F}{dt^2}=-e^{-tR}R[Q_0,R]e^{tR}+e^{-tR}
[Q_0,R]Re^{Rt}=e^{-tR}[[Q_0,R],R]]e^{tR} \ ,
\nonumber \\
\frac{d^nF}{dt^n}=e^{-tR}\overbrace{[[Q_0,R],\dots,],R]}^ne^{tR}
\end{eqnarray}
and consequently
\begin{equation}
Q_B=F(1)=Q_0+\sum_{n=1}^{\infty}
\frac{1}{n!}\overbrace{[[Q_0,R],R],\dots ]R]}^{n} \ .
\end{equation}
Then we obtain from  (\ref{A}) 
\begin{eqnarray}\label{A2}
Q_0(\Phi_0)+\sum_{n=2}^{\infty}
\frac{1}{n!}\overbrace{[[Q_0(\Phi_0),\Phi_0
],\dots,]\Phi_0]}^{n-1}=
\left(\sum_{n=1}^{\infty}
\frac{1}{n!}\overbrace{[[Q_B,R],\dots,],R]}^{n}\right)^L
\mathcal{I} \ . \nonumber \\
\end{eqnarray}
Now we must explain more carefully what we mean by the 
symbols $\mathcal{X}^{L,R}$ for any operator
$\mathcal{X}$.  Firstly, let us presume that  the
world-sheet  operator $\mathcal{X}$ is defined  by
\footnote{It is convenient to use "double trick" for open
strings. We trade the holomorphic and antiholomorphic
components of any field, defined in upper half plane, for
a single holomorphic field defined in the whole complex
plane.}
\begin{equation}
\mathcal{X}=\frac{1}{2\pi i} \oint_C dz x(z) \ ,
\end{equation}
where $C$ is any closed curve encircling the origin
of the conformal plane.  It
is clear that the operator given above can be written as
 a sum of two operators defined as integrals of
world-sheet densities over left half or right half of
the string, in particular for $Q,R$ we get 
\begin{equation}
Q_0=Q_0^R+Q_0^L, \ R=R^L+R^R \ ,
\end{equation}
where indices $R,L$ correspond to the left, right 
half of the open string respectively.
 For example, $Q_L$ can be defined
as a contour integral of the holomorphic density
over curve $C_L$ that lies in  the 
right half of the complex plane 
and $Q_R$ as a contour integral over curve $C_R$ 
that lies in the left half of the complex plane
\begin{equation}
Q_0^{R,L}=\frac{1}{2\pi i}\int_{C^R,C^L}dz j(z)_0 \ ,
R^{R,L}=\frac{1}{2\pi i} \int_{C^R,C^L}dz r(z) \ .
\end{equation}
 From  comments given above we immediately
obtain
\begin{equation}\label{QRRL}
[Q^R_0,R^L]=0
\end{equation}
since curves $C_R,C_L$  have not common points and
hence OPE between holomorphic densities is non-singular.
Then for any operator $\mathcal{O}$ that is commutator
of two  operators $\mathcal{X},\mathcal{Y}$ we get
\begin{equation}
\mathcal{O}=\mathcal{O}^R+
\mathcal{O}^L=[\mathcal{X},\mathcal{Y}]=
[\mathcal{X}^L,\mathcal{Y}^L]+
[\mathcal{X}^R,\mathcal{Y}^R]
\end{equation}
and consequently
\begin{equation}
\mathcal{O}^L=[\mathcal{X}^L,\mathcal{Y}^L] \ ,
\mathcal{O}^R=[\mathcal{X}^R,\mathcal{Y}^R] \ .
\end{equation}
Then we can write
\begin{equation}\label{Q0R}
[Q_0,R]^L\mathcal{I}=[Q_0^L,R^L]\mathcal{I} \    .
\end{equation}
Using previous results we claim that the solution of
the equation (\ref{A2}) has a form
\begin{equation}\label{phi0R}
\Phi_0=R^L\mathcal{I} \ .
\end{equation}
We will show that this expression really
leads to the BRST operator $Q_B$.
Firstly it is easy to see  that (\ref{phi0R}) solves
the following equation
\begin{equation}\label{Q0R1}
Q_0(\Phi_0)=[Q_0^L,R^L]\mathcal{I} \ 
\end{equation} 
since the left hand side of (\ref{Q0R1})  is equal to
\begin{eqnarray}
Q_0(\Phi_0)=(Q_0^R+Q_0^L)(R^L\mathcal{I})=
Q_0^RR^L\mathcal{I}+Q_0^LR^L\mathcal{I}=
R^LQ_0^R\mathcal{I}+Q_0^LR^L\mathcal{I}=
\nonumber \\
=-R^LQ_0^L\mathcal{I}+Q_0^LR^L\mathcal{I}=
[Q_0^L,R^L]\mathcal{I}
\nonumber \\
\end{eqnarray}
using (\ref{QRRL}) and (\ref{Qhor}).
   The second term in (\ref{A2})  is
\begin{equation}\label{A22}
[Q_0(\Phi_0),\Phi_0]=[[Q_0,R],R]^L\mathcal{I}=
[[Q_0,R]^L,R^L]\mathcal{I}=
[[Q_0^L,R^L],R^L]\mathcal{I} \ .
\end{equation}
In order to see that (\ref{phi0R}) really solves upper
expression and generally the whole equation
(\ref{A2}) we must perform lot of  calculations in
the similar way as 
in \cite{Horowitz}. Firstly, as  in case of $Q_0$ we can prove 
\begin{equation}
R(X)=R(X\star \mathcal{I})=R(X)\star \mathcal{I}+
X\star R(\mathcal{I}) \Rightarrow
R(\mathcal{I})=0 \ .
\end{equation}
This result follows from the contour integration argument.
Previous result implies
\begin{equation}\label{RL1}
R^L\mathcal{I}=-R^R\mathcal{I} \ .
\end{equation}
In the same way as in (\ref{Qhor}) we can
write
\begin{equation}\label{RL2}
R^R(X)\star Y=-X\star R^L (Y) \ .
\end{equation}
Note that there is not a factor $(-1)^X$ since $R$
is Grassmann even operator. 
Then we have
\begin{equation}
R^L\mathcal{I}\star R^L\mathcal{I}=
-R^R\mathcal{I}\star R^L\mathcal{I}=
\mathcal{I}\star R^LR^L(\mathcal{I})=
(R^L)^2(\mathcal{I}) \ ,
\end{equation}
where we have used in the first step (\ref{RL1}) and
in the second  step (\ref{RL2}) in the form
\begin{equation}
R^R(\mathcal{I})\star R^L(\mathcal{I})=
-\mathcal{I}\star R^L(R^L(\mathcal{I})) \ .
\end{equation}
 We also have
\begin{eqnarray}
Q^R_0(R^L\mathcal{I})\star \mathcal{I}=
-R^L\mathcal{I}\star Q^L_0\mathcal{I}\Rightarrow
R^LQ^R_0(\mathcal{I})\star\mathcal{I}=
-R^L\mathcal{I}\star Q^L_0(\mathcal{I})\Rightarrow\nonumber \\
\Rightarrow R^LQ^L_0(\mathcal{I})=R^L(\mathcal{I})
\star Q^L_0(\mathcal{I}) \ ,
\end{eqnarray}
where  we 
have firstly used (\ref{QRRL}) and then
 (\ref{Qhor}).
 In the same way we can show that
\begin{eqnarray}
R^R(Q_0^L \mathcal{I})\star\mathcal{I}=
-Q_0^L(\mathcal{I})\star R^L\mathcal{I}\Rightarrow
Q_0^LR^R\mathcal{I}=-Q_0^L\mathcal{I}\star R^L\mathcal{I}
\Rightarrow 
Q^L_0(R^L\mathcal{I})=Q^L_0\mathcal{I}\star R^L\mathcal{I}
 \ ,\nonumber \\
-Q_0^R((R^L)^2\mathcal{I})\star\mathcal{I}=
(R^L)^2\mathcal{I}\star Q_0^L\mathcal{I}
\Rightarrow (R^L)^2 Q_0^L\mathcal{I}=
(R^L)^2\mathcal{I}\star Q^L\mathcal{I} \ , \nonumber \\
R^R(Q_0^L(R^L\mathcal{I}))\star\mathcal{I}=
-Q_0^L(R^L\mathcal{I})\star R^L\mathcal{I}
\Rightarrow \nonumber \\
\Rightarrow 
Q_0^L((R^L)^2\mathcal{I})=Q_0^L(R^L\mathcal{I})\star
R^L\mathcal{I} \Rightarrow
Q_0^L(R^L)^2\mathcal{I}=Q_0^L\mathcal{I}\star (R^L)^2
\mathcal{I} \ . \nonumber \\
\end{eqnarray}
Generally we obtain
\begin{eqnarray}
(R^L)^nQ_0^L\mathcal{I}=(R^L)^n\mathcal{I}
\star Q_0^L\mathcal{I} \ , \nonumber \\
Q_0^L(R^L)^n\mathcal{I}=Q_0^L\mathcal{I}
\star (R^L)^n\mathcal{I} \ . \nonumber \\
\end{eqnarray}
We can also write
\begin{eqnarray}
(R^L)^nQ_0^L \mathcal{I}\star R^L\mathcal{I}=
-R^R((R^L)^nQ_0^L\mathcal{I})\Rightarrow \nonumber \\
\Rightarrow 
(R^L)^nQ_0^L\mathcal{I}\star R^L\mathcal{I}=
-(R^L)^nQ_0^LR^R\mathcal{I} \Rightarrow
(R^L)^nQ_0^L\mathcal{I}\star R^L\mathcal{I}=
(R^L)^nQ_0^LR^L\mathcal{I}  ; \ \nonumber \\
R^L\mathcal{I}\star Q_0^L(R^L)^n\mathcal{I}=
-R^R\mathcal{I}\star Q_0^n(R^L)^n\mathcal{I}
=\mathcal{I}\star R^L(Q_0^L (R^L)^n \mathcal{I})=
R^LQ_0^L(R^L)^n\mathcal{I} \ . \nonumber \\
\end{eqnarray}
Generally we have
\begin{eqnarray}
R^L\mathcal{I}\star (R^L)^nQ_0^L(R^L)^m\mathcal{I}=
(R^L)^{n+1}Q_0^L(R^L)^m\mathcal{I} ; \ \nonumber \\ 
(R^L)^nQ_0^L(R^L)^m\mathcal{I}\star R^L\mathcal{I}=
(R^L)^nQ_0^L(R^L)^{m+1}\mathcal{I} \ . \nonumber \\
\end{eqnarray}
Using these results we can easily calculate
\begin{eqnarray}
[Q_0(\Phi_0),\Phi_0]
=[[Q_0^L,R^L]\mathcal{I},R^L\mathcal{I}]=
Q_0^L\mathcal{I}\star R^L\mathcal{I}
\star R^L\mathcal{I}-R^L\mathcal{I}\star Q_0^LR^L
\mathcal{I}-\nonumber \\
-R^L\mathcal{I}\star 
Q^L_0R^L\mathcal{I}
+R^L\mathcal{I}\star R^L\mathcal{I}\star Q_0^L\mathcal{I}=
\nonumber \\=
(Q_0^L(R^L)^2-R^LQ_0^LR^L-R^LQ^L_0R^L+
(R^L)^2Q^L_0)\mathcal{I}=
[[Q_0^L,R^L],R^L]\mathcal{I} \  \nonumber \\
\end{eqnarray}
which proves (\ref{A22}). 
 Generally we have a result
\begin{equation}
\overbrace{[[Q_0(\Phi_0),\Phi_0],\dots],\Phi_0]}^{n-1}=
\overbrace{[[Q_0,R],R\dots,],R]^L}^{n}\mathcal{I} \ .
\end{equation}
We can prove upper relation by mathematical
induction. 
It was shown that this relation holds for $n=1,2$.
Let us presume its validity for $n=N-1$. Then 
we have for $n=N$
\begin{eqnarray}
\overbrace{[[[Q_0(\Phi_0),\Phi_0],\dots],\Phi_0]}^{N-1}=
[\overbrace{[[Q_0,R],R,\dots,]R]^L}^{N-1}\mathcal{I},\Phi_0]=
\nonumber \\
=\overbrace{[[Q_0,R],R],\dots],R]^L}^{N-1}\mathcal{I}\star R^L\mathcal{I}-
R^L\mathcal{I}\star \overbrace{[[Q_0,R],R],\dots],R]^L}^{N-1}\mathcal{I}
=\nonumber \\
=\overbrace{[[Q_0,R],R],\dots],R]^L}^{N-1}R\mathcal{I}-
R\overbrace{[[Q_0,R],R],\dots],R]^L}^{N-1}\mathcal{I}=
\overbrace{[[Q_0,R],\dots],R]^L}^{N}\mathcal{I} \ 
\nonumber \\
\end{eqnarray}
using results given above. 
We can then claim that $\Phi_0=R^L\mathcal{I}$ solves
(\ref{A2}). 
We have also shown above that $A=e^{\Phi_0}Q_B(e^{-\Phi_0})$
solves the equation of motion (\ref{eqm1}). In other words,
we have found  classical field  in the  background
independent NSFT that leads to the NSFT with correct
BRST operator corresponding to some particular
CFT background. It  can be shown in the same way
as in \cite{Horowitz}  that for some
particular background CFT characterised by
 $Q_B$ there is unique field $\Phi_0=
R^L\mathcal{I}$. 

In this section we have proposed background independent formulation
(at least formally) 
of the NSFT. We have shown that any CFT background arises
as a particular solution of the background independent NSFT theory.
Of course, our calculation was  pure formal as in case of
\cite{Horowitz} so that more detailed analysis should be done.
We  return to some open questions and suggestions for further work
in conclusion.

\section{Conclusion}\label{fifth}

In this short note we have  proposed   a background
independent formulation of Berkovits' string field theory
\cite{Berkovits1,Berkovits2,Berkovits3}. Our proposal is
based on the form of the BRST operator
for RNS string theory presented in \cite{BerkovitsBRST}.
Since the BRST operator we started in the background independent
NSFT with contains ghost  fields only, 
it is background independent as well. This is the
situation similar to the case of vacuum string field
theory 
\cite{SenV1,SenV2,SenV3,SenV4}, where the
BRST operator is constructed from the ghost field
only and consequently VSFT is background independent.
On the other hand, the BRST operator in VSFT has
a  vanishing cohomology, while in NSFT it has not.
We mean that this is not in contradiction since our
background independent NSFT theory corresponds to
the BPS object so that there is no  place for  tachyon condensation.
This remark also suggests possible limitation of
our proposal. We mean that the true background independent
formulation of supersymmetric string field  theory will have such a form
that will allow any solutions corresponding to BPS or
non-BPS CFT background. In order to find such a
formulation, we think that complete supersymmetric invariant
action, including Ramond sector, should be found.
We believe that such a formulation will be found in the
near future and it will allow us to get new insight
in the nature of supersymmetric theory.

We must also stress one important thing. Our calculation 
was pure formal in the sense  that we have worked with Witten's
star product as with an abstract object that 
does not depend on any background. 
It would be nice to perform more detailed analysis based
on the CFT technique, in the similar way as in
the beautiful paper  \cite{SenV4}. We hope to
return to this problem in the future. 
\\
\\
{\bf Acknowledgements}
We would like to thank N. Berkovits for useful comments
and for pointing out the work \cite{BerkovitsBRST}. 
 This work was supported by the
Czech Ministry of Education under Contract No.
1443100006.


\end{document}